\documentclass{emulateapj}
\usepackage{apjfonts}
\usepackage{rotating}
\usepackage{enumerate}

\newcommand{\pivec}{\mbox{\boldmath $\pi$}}
\newcommand{\muvec}{\mbox{\boldmath $\mu$}}

\lefthead{JEONG ET AL.} 
\righthead{Reanalyses of Events in OGLE EWS Database
}

\begin{document}
\title{Reanalyses of Anomalous Gravitational Microlensing Events in the OGLE-III Early Warning System Database with Combined Data
}

\author{
J. Jeong\altaffilmark{1},
H. Park\altaffilmark{1},
C. Han\altaffilmark{1,U,$\dagger$},
A. Gould\altaffilmark{U2,U}, 
A. Udalski\altaffilmark{O1,O}, \\
and\\
M. K. Szyma{\'n}ski\altaffilmark{O1}, 
G. Pietrzy{\'n}ski\altaffilmark{O1}, 
I. Soszy{\'n}ski\altaffilmark{O1}, 
R. Poleski\altaffilmark{O1,U2},
K. Ulaczyk\altaffilmark{O1},
{\L}. Wyrzykowski\altaffilmark{O1,O2},\\
(The OGLE Collaboration),\\
F. Abe\altaffilmark{M2},
D. P. Bennett\altaffilmark{M11,P},
I. A. Bond\altaffilmark{M3},
C. S. Botzler\altaffilmark{M4},
M. Freeman\altaffilmark{M4},
A. Fukui\altaffilmark{M6},
D. Fukunaga\altaffilmark{M2},
Y. Itow\altaffilmark{M2},
N. Koshimoto\altaffilmark{M1},
K. Masuda\altaffilmark{M2},
Y. Matsubara\altaffilmark{M2},
Y. Muraki\altaffilmark{M2},
S. Namba\altaffilmark{M1},
K. Ohnishi\altaffilmark{M7},
N. J. Rattenbury\altaffilmark{M4},
To. Saito\altaffilmark{M8},
D. J. Sullivan\altaffilmark{M5},
W. L. Sweatman\altaffilmark{M3},
T. Sumi\altaffilmark{M2},
D. Suzuki\altaffilmark{M1},
P. J. Tristram\altaffilmark{M9},
N. Tsurumi\altaffilmark{M2},
K. Wada\altaffilmark{M1},
N. Yamai\altaffilmark{M10},
P. C. M. Yock\altaffilmark{M4},
A. Yonehara\altaffilmark{M10},\\
(The MOA Collaboration),\\
M. D. Albrow\altaffilmark{P1},
V. Batista\altaffilmark{U2},
J.-P. Beaulieu\altaffilmark{P3},
J. A. R. Caldwell\altaffilmark{P4},
A. Cassan\altaffilmark{P3},
A. Cole\altaffilmark{P4},
C. Coutures\altaffilmark{P3},
S. Dieters\altaffilmark{P3},
M. Dominik\altaffilmark{P6,R},
D. Dominis Prester\altaffilmark{P7},
J. Donatowicz\altaffilmark{P8},
P. Fouqu\'e\altaffilmark{P9,P17},
J. Greenhill\altaffilmark{P5},
M. Hoffman\altaffilmark{P10},
M. Huber\altaffilmark{P11},
U. G. J{\o}rgensen\altaffilmark{P12},
S. R. Kane\altaffilmark{P13},
D. Kubas\altaffilmark{P3},
R. Martin\altaffilmark{P14},
J.-B. Marquette\altaffilmark{P3},
J. Menzies\altaffilmark{P15},
C. Pitrou\altaffilmark{P3},
K. Pollard\altaffilmark{P1},
K. C. Sahu\altaffilmark{P16},
C. Vinter\altaffilmark{P12},
J. Wambsganss\altaffilmark{P18},
A. Williams\altaffilmark{P14},\\
(The PLANET Collaboration),\\
W. Allen\altaffilmark{U3},
G. Bolt\altaffilmark{U4},
J.-Y. Choi\altaffilmark{1},
G. W. Christie\altaffilmark{U5},
D. L. DePoy\altaffilmark{U6},
J. Drummond\altaffilmark{U7},
B. S. Gaudi\altaffilmark{U2},
K.-H. Hwang\altaffilmark{1},
Y. K. Jung\altaffilmark{1},
C.-U. Lee\altaffilmark{U11},
F. Mallia\altaffilmark{U12},
D. Maoz\altaffilmark{U9},
A. Maury\altaffilmark{U12},
J. McCormick\altaffilmark{U13},
L. A. G. Monard\altaffilmark{U14},
D. Moorhouse\altaffilmark{U15},
T. Natusch\altaffilmark{U5,U16},
E. O. Ofek\altaffilmark{U17},
B.-G. Park\altaffilmark{U11},
R. W. Pogge\altaffilmark{U2},
R. Santallo\altaffilmark{U18},
I.-G. Shin\altaffilmark{1},
G. Thornley\altaffilmark{U15},
J. C. Yee\altaffilmark{U2,U19,SF},\\
(The $\mu$FUN Collaboration),\\
A. Allan\altaffilmark{R1},
D. M. Bramich\altaffilmark{R2},
M. J. Burgdorf\altaffilmark{R3}, 
K. Horne\altaffilmark{P6,P},
M. Hundertmark\altaffilmark{P6,P12},
N. Kains\altaffilmark{R5,P6},
C. Snodgrass\altaffilmark{R7},
I. Steele\altaffilmark{R8},
R. Street\altaffilmark{R9,P},
Y. Tsapras\altaffilmark{R9,R10,P}\\
(The RoboNet Collaboration)
}

\affil{$^{1}$Department of Physics, Institute for Astrophysics, Chungbuk National University, Cheongju 361-763, Korea}
\affil{$^{O1}$Warsaw University Observatory, Al.~Ujazdowskie~4, 00-478~Warszawa, Poland}
\affil{$^{O2}$Institute of Astronomy, University of Cambridge, Madingley Road, Cambridge CB3 oHA, UK}
\affil{$^{M1}$Department of Earth and Space Science, Osaka University, Osaka 560-0043, Japan}
\affil{$^{M2}$Solar-Terrestrial Environment Laboratory, Nagoya University, Nagoya, 464-8601, Japan}
\affil{$^{M3}$Institute of Information and Mathematical Sciences, Massey University, Private Bag 102-904, North Shore Mail Centre, Auckland, New Zealand}
\affil{$^{M4}$Department of Physics, University of Auckland, Private Bag 92-019, Auckland 1001, New Zealand}
\affil{$^{M5}$School of Chemical and Physical Sciences, Victoria University, Wellington, New Zealand}
\affil{$^{M6}$Okayama Astrophysical Observatory, National Astronomical Observatory of Japan, Asakuchi, Okayama 719-0232, Japan}
\affil{$^{M7}$Nagano National College of Technology, Nagano 381-8550, Japan}
\affil{$^{M8}$Tokyo Metropolitan College of Aeronautics, Tokyo 116-8523, Japan}
\affil{$^{M9}$Mt. John University Observatory, P.O. Box 56, Lake Tekapo 8770, New Zealand}
\affil{$^{M10}$Department of Physics, Faculty of Science, Kyoto Sangyo University, 603-8555, Kyoto, Japan}
\affil{$^{M11}$Department of Physics, University of Notre Dame, 225 Nieuwland Science Hall, Notre Dame, IN 46556-5670, USA}
\affil{$^{P1}$University of Canterbury, Department of Physics and Astronomy, Private bag 4800, Christchurch 8020, New Zealand}
\affil{$^{P3}$Institut d'Astrophysique de Paris, Universit'e Pierre et Marie Curie, CNRS UMR7095, 98bis Boulevard Arago, 75014 Paris, France}
\affil{$^{P4}$McDonald Observatory, 16120 St Hwy Spur 78 \#2, Fort Davis, TX 79734 USA}
\affil{$^{P5}$University of Tasmania, School of Mathematics and Physics, Private Bag 37, GPO Hobart, Tas 7001, Australia}
\affil{$^{P6}$SUPA, University of St Andrews, School of Physics \& Astronomy, North Haugh, St Andrews, KY16 9SS, United Kingdom}
\affil{$^{P7}$Department of Physics, Faculty of Arts and Sciences, University of Rijeka, Omladinska 14, 51000 Rijeka, Croatia}
\affil{$^{P8}$Department of Computing, Technical University of Vienna, Wiedner Hauptstrasse 10, A-1040 Vienna, Austria}
\affil{$^{P9}$IRAP, UMR 5277, CNRS, Universit\'e de Toulouse, 14 av. E. Belin, 31400 Toulouse, France}
\affil{$^{P10}$University of the Free State, Faculty of Natural and Agricultural Sciences, Department of Physics, PO Box 339, Bloemfontein 9300, South Africa}
\affil{$^{P11}$Institute for Astronomy, University of Hawaii, 2680 Woodlawn Drive Honolulu, HI 96822-1839, USA}
\affil{$^{P12}$Niels Bohr Institute, University of Copenhagen, Juliane Maries vej 30, 2100 Copenhagen, Denmark}
\affil{$^{P13}$Department of Physics and Astronomy, San Francisco State University, 1600 Holloway Avenue, San Francisco, CA 94132, USA}
\affil{$^{P14}$Perth Observatory, Walnut Road, Bickley, Perth 6076, WA, Australia}
\affil{$^{P15}$South African Astronomical Observatory, PO box 9, Observatory 7935, South Africa}
\affil{$^{P16}$Space Telescope Science Institute, 3700 San Martin Drive, Baltimore, MD 21218, USA}
\affil{$^{P17}$CFHT Corporation, 65-1238 Mamalahoa Hwy, Kamuela, HI, 96743, USA}
\affil{$^{P18}$Astronomisches Rechen-Institut, M{\"o}nchhofstr. 12-14, 69120 Heidelberg, Germany}
\affil{$^{U2}$Department of Astronomy, Ohio State University, 140 W. 18th Ave., Columbus, OH 43210, USA}
\affil{$^{U3}$Vintage Lane Observatory, Blenheim, New Zealand}
\affil{$^{U4}$Craigie Observatory, Craigie, Western Australia}
\affil{$^{U5}$Auckland Observatory, Auckland, New Zealand}
\affil{$^{U6}$Department of Physics and Astronomy, Texas A\&M University, College Station, TX 77843, USA}
\affil{$^{U7}$Possum Observatory, Patutahi, Gisbourne, New Zealand}
\affil{$^{U9}$School of Physics and Astronomy and Wise Observatory, Tel-Aviv University, Tel-Aviv 69978, Israel}
\affil{$^{U11}$Korea Astronomy and Space Science Institute, Daejon 305-348, Korea}
\affil{$^{U12}$Campo Catino Austral Observatory, San Pedro de Atacama, Chile}
\affil{$^{U13}$Farm Cove Observatory, Centre for Backyard Astrophysics, Pakuranga, Auckland, New Zealand}
\affil{$^{U14}$Klein Karoo Observatory, Calitzdorp, and Bronberg Observatory, Pretoria, South Africa}
\affil{$^{U15}$Kumeu Observatory, Kumeu, New Zealand}
\affil{$^{U16}$Institute for Radiophysics and Space Research, AUT University, Auckland, New Zealand}
\affil{$^{U17}$Department of Particle Physics \& Astrophysics, Weizmann Institute of Science, Rehovot 76100, Israel}
\affil{$^{U18}$Southern Stars Observatory, Faaa, Tahiti, French Polynesia}
\affil{$^{U19}$Harvard-Smithsonian Center for Astrophysics, 60 Garden St, Cambridge, MA 02138, USA}
\affil{$^{R1}$School of Physics, University of Exeter, Stocker Road, Exeter EX4 4QL, UK}
\affil{$^{R2}$Qatar Environment and Energy Research Institute, Qatar Foundation, PO Box 5825, Doha, Qatar}
\affil{$^{R3}$HE Space Operations, Flughafenallee 24, 28199 Bremen, Germany}
\affil{$^{R5}$Space Telescope Science Institute, 3700 San Martin Drive, Baltimore, MD 21218, USA}
\affil{$^{R7}$Planetary and Space Sciences, Department of Physical Sciences, The Open University, Milton Keynes, MK7 6AA, UK}
\affil{$^{R8}$Astrophysics Research Institute, Liverpool John Moores University, Liverpool CH41 1LD, UK}
\affil{$^{R9}$Las Cumbres Observatory Global Telescope Network, 6740 Cortona Drive, Suite 102, Goleta, CA 93117, USA}
\affil{$^{R10}$Astronomisches Rechen-Institut, Zentrum f{\"u}r Astronomie der Universit{\"a}t Heidelberg (ZAH), 69120 Heidelberg, Germany}

\footnotetext[O]{The OGLE Collaboration.}
\footnotetext[M]{The MOA Collaboration.}
\footnotetext[P]{The PLANET Collaboration.}
\footnotetext[U]{The $\mu$FUN Collaboration.}
\footnotetext[R]{The RoboNet Collaboration.}
\footnotetext[$\dagger$]{Corresponding author.}
\footnotetext[SF]{Sagan Fellow.}

\begin{abstract}

We reanalyze microlensing events in the published list of anomalous events that were observed from the OGLE lensing survey conducted during 2004-2008 period. In order to check the existence of possible degenerate solutions and extract extra information, we conduct analyses based on combined data from other survey and follow-up observation and consider higher-order effects. Among the analyzed events, we present analyses of 8 events for which either new solutions are identified or additional information is obtained. We find that the previous binary-source interpretations of 5 events are better interpreted by binary-lens models. These events include OGLE-2006-BLG-238, OGLE-2007-BLG-159, OGLE-2007-BLG-491, OGLE-2008-BLG-143, and OGLE-2008-BLG-210. With additional data covering caustic crossings, we detect finite-source effects for 6 events including OGLE-2006-BLG-215, OGLE-2006-BLG-238, OGLE-2006-BLG-450, OGLE-2008-BLG-143, OGLE-2008-BLG-210, and OGLE-2008-BLG-513. Among them, we are able to measure the Einstein radii of 3 events for which multi-band data are available. These events are OGLE-2006-BLG-238, OGLE-2008-BLG-210, and OGLE-2008-BLG-513. For OGLE-2008-BLG-143, we detect higher-order effect induced by the changes of the observer's position caused by the orbital motion of the Earth around the Sun. In addition, we present degenerate solutions resulting from the known close/wide or ecliptic degeneracy. Finally, we note that the masses of the binary companions of the lenses of OGLE-2006-BLG-450 and OGLE-2008-BLG-210 are in the brown-dwarf regime.
\end{abstract}


\section{Introduction}
Light curves of gravitational microlensing events are characterized by a smooth, symmetric, and non-repeating shape \citep{paczynski86}. However, they often exhibit deviations from the standard form. The causes of the deviations include the binarity of either lensing objects \citep[binary-lens events:][]{mao91} or lensed source stars \citep[binary-source events:][]{griest92}. Deviations can also arise due to higher-order effects caused by the finite size of a source star \citep[finite-source effect:][]{nemiroff94} and the positional changes of the observer \citep[parallax effect:][]{gould92}, lens \citep[lens-orbital effect:][]{albrow00,park13}, and source star \citep[xallarap effect:][]{alcock01} induced by the orbital motions of the observer (Earth), lens, and source star, respectively. 

Analyzing anomalies in lensing light curves is important because it provides useful information about the lenses and lensed stars. By analyzing the light curve of a binary-lens event, one can obtain information about the mass ratio between the lens components. Analyzing the light curve of a binary-source event yields the flux ratio between the binary-source components. If finite-source and parallax effects are simultaneously detected, one can uniquely determine the physical parameters of the lens including the mass and distance to the lens \citep{gould92}. 

However, interpretation of anomalous lensing events is difficult due to various reasons. One important reason is the degeneracy problem where lensing solutions based on different interpretations result in a similar anomaly pattern. For example, it is known that binary-lens and binary-source solutions can often result in a similar anomaly  pattern \citep{gaudi98}. Even if an anomaly is identified to be caused by a binary lens, $\chi^2$ distributions in the space of the lensing parameters are usually complex due to the nonlinear dependence of lensing magnifications on the parameters and thus there often exist multiple local minima resulting in similar anomaly patterns. 

For the correct interpretation of lensing event by resolving the degeneracy problem, dense and continuous coverage of lensing light curves is important. Good coverage data are also important in extracting extra information about lenses and lensed stars by detecting subtle deviations caused by higher-order effects. In order to obtain correct interpretations of events and maximize information about lens systems, therefore, it is important to test all possible causes of anomalies based on all available data. 
\begin{deluxetable*}{lrrrr}
\centering
\tablecaption{Events and Coordinates\label{table:one}}
\tablewidth{0pt}
\tablehead{
\multicolumn{1}{c}{Event} &
\multicolumn{1}{c}{RA (J2000)} &
\multicolumn{1}{c}{DEC (J2000)} &
\multicolumn{1}{c}{$\lambda$} &
\multicolumn{1}{c}{$\beta$} 
}
\startdata
OGLE-2006-BLG-215                   & 17$^{\rm h}$59$^{\rm m}$06$^{\rm s}$\hskip-2pt.63 & -29$^\circ$08${\rm '}$55${\rm ''}$\hskip-2pt.7 & 269$^\circ$\hskip-2pt.80 & -5$^\circ$\hskip-2pt.71 \\
OGLE-2006-BLG-238/MOA-2006-BLG-26   & 17$^{\rm h}$57$^{\rm m}$11$^{\rm s}$\hskip-2pt.66 & -30$^\circ$48${\rm '}$06${\rm ''}$\hskip-2pt.0 & 269$^\circ$\hskip-2pt.39 & -7$^\circ$\hskip-2pt.36 \\
OGLE-2006-BLG-450                   & 17$^{\rm h}$52$^{\rm m}$00$^{\rm s}$\hskip-2pt.47 & -31$^\circ$07${\rm '}$36${\rm ''}$\hskip-2pt.3 & 268$^\circ$\hskip-2pt.27 & -7$^\circ$\hskip-2pt.70 \\
OGLE-2007-BLG-159                   & 17$^{\rm h}$54$^{\rm m}$41$^{\rm s}$\hskip-2pt.68 & -29$^\circ$25${\rm '}$39${\rm ''}$\hskip-2pt.5 & 268$^\circ$\hskip-2pt.84 & -5$^\circ$\hskip-2pt.99 \\
OGLE-2007-BLG-491                   & 17$^{\rm h}$56$^{\rm m}$28$^{\rm s}$\hskip-2pt.68 & -32$^\circ$13${\rm '}$15${\rm ''}$\hskip-2pt.3 & 269$^\circ$\hskip-2pt.25 & -8$^\circ$\hskip-2pt.78 \\
OGLE-2008-BLG-143/MOA-2008-BLG-111  & 17$^{\rm h}$59$^{\rm m}$38$^{\rm s}$\hskip-2pt.58 & -28$^\circ$41${\rm '}$34${\rm ''}$\hskip-2pt.8 & 269$^\circ$\hskip-2pt.92 & -5$^\circ$\hskip-2pt.25 \\
OGLE-2008-BLG-210/MOA-2008-BLG-177  & 17$^{\rm h}$59$^{\rm m}$22$^{\rm s}$\hskip-2pt.97 & -27$^\circ$42${\rm '}$30${\rm ''}$\hskip-2pt.2 & 269$^\circ$\hskip-2pt.86 & -4$^\circ$\hskip-2pt.27 \\
OGLE-2008-BLG-513/MOA-2008-BLG-401  & 17$^{\rm h}$52$^{\rm m}$43$^{\rm s}$\hskip-2pt.51 & -30$^\circ$51${\rm '}$33${\rm ''}$\hskip-2pt.2 & 268$^\circ$\hskip-2pt.43 & -7$^\circ$\hskip-2pt.43 
\enddata
\end{deluxetable*}

\begin{deluxetable*}{ll}
\tablecaption{Telescopes\label{table:two}}
\tablewidth{0pt}
\tablehead{
\multicolumn{1}{c}{Event} &
\multicolumn{1}{c}{Telescopes} 
}
\startdata
OGLE-2006-BLG-215                  & OGLE ({\it I}), CTIO ({\it I}), LOAO ({\it I}), Auckland ({\it N}), Danish ({\it I}), Boyden ({\it I})\\
OGLE-2006-BLG-238/MOA-2006-BLG-26  & OGLE ({\it I}), MOA ({\it R}), CTIO ({\it I}), CTIO ({\it V}), LOAO ({\it I}), Wise ({\it N}), SAAO ({\it I}), Danish ({\it I}), Danish ({\it V}),\\
                                   & Canopus ({\it I}), Boyden ({\it I}), Perth ({\it I}), FTN ({\it R}), LT ({\it R})\\
OGLE-2006-BLG-450                  & OGLE ({\it I}), SAAO ({\it I}), Danish ({\it I}), Canopus ({\it I}), LT ({\it R})\\
OGLE-2007-BLG-159                  & OGLE ({\it I}), CTIO ({\it I}), Auckland ({\it R}), Canopus ({\it I}), FCO ({\it N}), Perth ({\it I}), Boyden ({\it I}), Danish ({\it I}), VLO ({\it N})\\
OGLE-2007-BLG-491                  & OGLE ({\it I}) \\
OGLE-2008-BLG-143/MOA-2008-BLG-111 & OGLE ({\it I}), MOA ({\it R})\\
OGLE-2008-BLG-210/MOA-2008-BLG-177 & OGLE ({\it I}), MOA ({\it R}), CTIO ({\it I}), CTIO ({\it V}), Wise ({\it R}), Bronberg ({\it N}), Canopus ({\it I}), Perth ({\it I}), SAAO ({\it I}),\\
                                   & FTN ({\it R}), FTS ({\it R})\\
OGLE-2008-BLG-513/MOA-2008-BLG-401 & OGLE ({\it I}), MOA ({\it R}), CTIO ({\it I}), CTIO ({\it V}), Kumeu ({\it I}), Palomar ({\it I}), Craigie ({\it N}), VLO ({\it N}), Bronberg ({\it N}),\\
                                   & Auckland ({\it R}), CAO ({\it N}), FCO ({\it N}), Possum ({\it N}), SSO ({\it N}), Wise ({\it R}), Canopus ({\it I}), Perth ({\it I}), FTN ({\it R}),\\
                                   & FTS ({\it R}), LT ({\it R})
\enddata
\tablecomments{OGLE: Warsaw 1.3m, Las Campanas Observatory, Chile; MOA: 1.8m, Mt. John Observatory, New Zealand; CTIO ($\mu$FUN): 1.3m, Cerro Tololo Inter-American Observatory, Chile; Auckland ($\mu$FUN): 0.40m, Auckland Observatory, New Zealand; FCO ($\mu$FUN): 0.36m, Farm Cove Observatory, New Zealand; VLO ($\mu$FUN): 0.4m, Vintage Lane Observatory, New Zealand; Bronberg ($\mu$FUN): 0.3m, Bronberg Observatory, South Africa; Kumeu ($\mu$FUN): 0.36m, Kumeu Observatory, New Zealand; Possum ($\mu$FUN): 0.36m, Possum Observatory, New Zealand; Palomar ($\mu$FUN): 1.5m, Palomar Observatory, California, USA; LOAO ($\mu$FUN): 1.0m, Mt. Lemmon Observatory, Tucson, Arizona, USA; Wise ($\mu$FUN): 1.0m, Wise Observatory, Israel; Perth ($\mu$FUN): 0.6m, Perth Observatory, Australia; Boyden (PLANET): 1.5m, Boyden Observatory, South Africa; SAAO (PLANET): 1.0m, South African Astronomical Observatory, South Africa; Canopus (PLANET): 1.0m, Canopus Hill Observatory, Tasmania, Australia; FTN (RoboNet): 2.0m, Faulkes North, Hawaii; FTS (RoboNet): 2.0m, Faulkes South, Australia; LT (RoboNet): 2.0m, Liverpool Telescope, LaPalma, Spain; Danish (PLANET): 1.54m Danish Telescope, European Southern Observatory, La Silla, Chile. The notation in the parentheses after each telescope denotes the filter used for observation. MOA-red band is a custom wide band where the band width roughly corresponds to the sum of {\it R} and {\it I} band. The filter notation ``{\it N}'' denotes that no filter is used. 
}
\end{deluxetable*}

In a series of papers \citep{jaroszynski06,jaroszynski10,skowron07}, the Optical Gravitational Lensing Experiment \citep[OGLE:][]{udalski03} group published lists of anomalous lensing events in the OGLE-III Early Warning System (EWS) database that is collected from the lensing survey conducted during 2004 -- 2008 period (hereafter the ``OGLE Anomaly Catalog''). They also presented solutions of the anomalous events based on binary-lens and binary-source interpretations. 

In this work, we reanalyze the anomalous events in the OGLE Anomaly Catalog. We conduct thorough search for local solutions in order to investigate possible degeneracy. For analyses based on better coverage of anomalies, we include additional data obtained from other survey and follow-up observations. In addition, we consider higher-order effects that were not considered in the previous analyses.

\section{Combined Data}
Among the total 68 events in the OGLE Anomaly Catalog, we conduct reanalyses of events for which the anomalies and overall light curves are well covered either by the OGLE data alone or with the addition of data from other survey and follow-up observations. Among the analyzed events, we present the results of 8 events for which either new solutions are identified (5 events) or additional information of the Einstein radius (3 events) or the lens parallax (1 event) is obtained. 

Table~\ref{table:one} shows the list of events analyzed in this work along with their equatorial and ecliptic coordinates. Table~\ref{table:two} shows the data sets used in our analyses and the telescopes used for observation. Except for the event OGLE-2007-BLG-491, we use extra data in addition to the data from the OGLE survey. These additional data were taken from the survey conducted by the Microlensing Observations in Astrophysics \citep[MOA:][]{bond01,sumi03} group and the follow-up observation conducted by the Microlensing Follow-Up Network \citep[$\mu$FUN:][]{gould06}, Probing Lensing Anomalies NETwork \citep[PLANET:][]{beaulieu06}, and RoboNet \citep{tsapras09}. We note that the OGLE data used in the OGLE Anomaly Catalog were based on the online data processed by automatic pipeline. In this work, we used new sets of data prepared by conducting rereduction of the data that is optimized for the individual events.

\begin{figure}[ht]
\epsscale{1.08}
\plotone{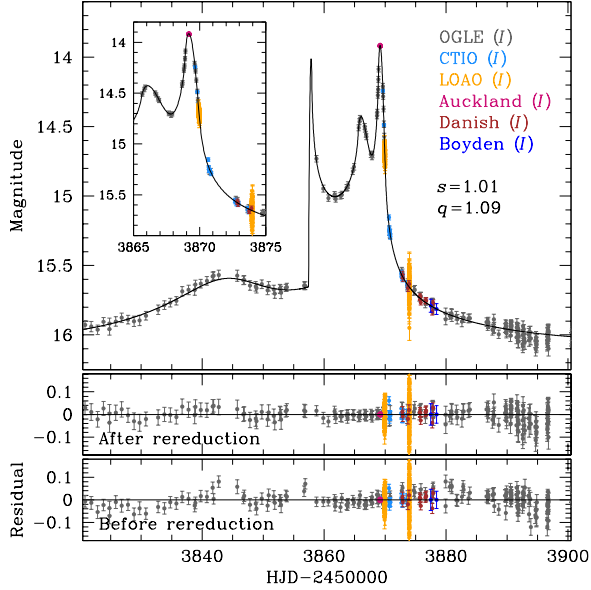}
\caption{\label{fig:one}
Light curve of OGLE-2006-BLG-215. Solid line is the best-fit model from our analysis. The two lower panels show the residuals 
of data sets before and after optimized re-reduction.
}\end{figure}

\begin{figure}[ht]
\epsscale{1.08}
\plotone{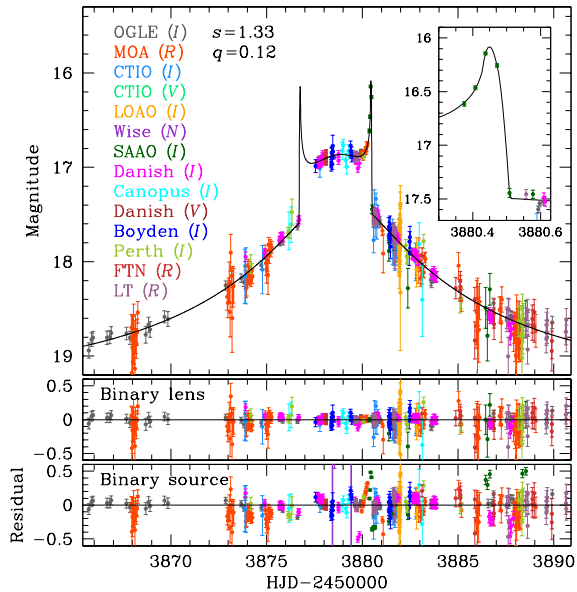}
\caption{\label{fig:two}
Light curve of OGLE-2006-BLG-238. The two lower panels show the residuals from the binary-lens and binary-source models. With the coverage of the caustic exit by the additional data, finite source-effects are detected.
}\end{figure}

\begin{figure}[ht]
\epsscale{1.08}
\plotone{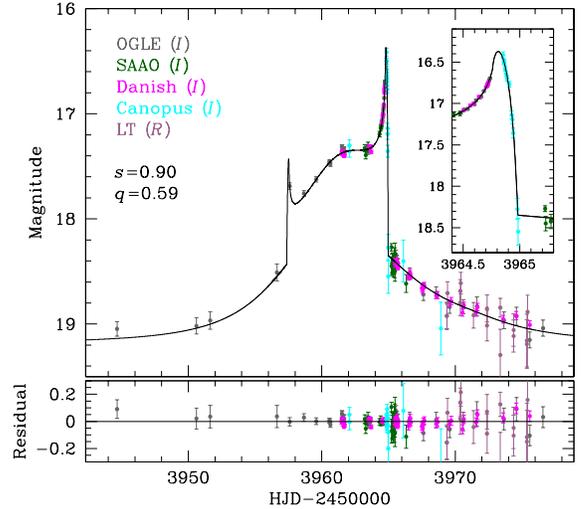}
\caption{\label{fig:three}
Light curve of OGLE-2006-BLG-450. With additional data covering the caustic exit, finite-source effects are detected.
}\end{figure}

\begin{figure}[ht]
\epsscale{1.08}
\plotone{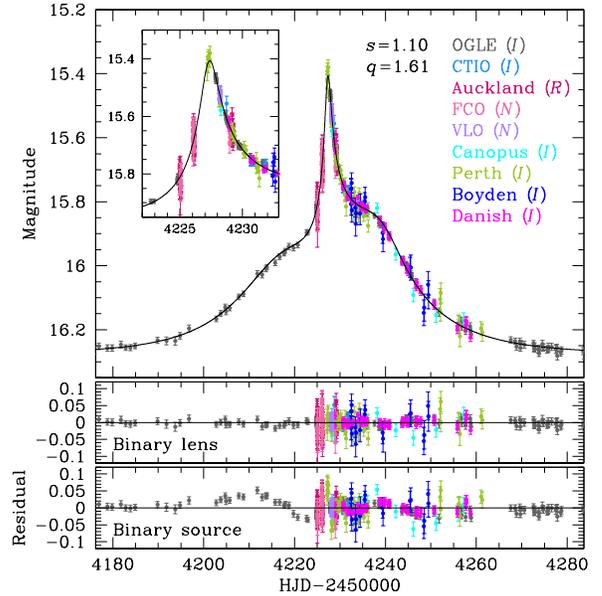}
\caption{\label{fig:four}
Light curve of OGLE-2007-BLG-159. The two lower panels show the residuals from the binary-lens and binary-source models.
}\end{figure}

\begin{figure}[ht]
\epsscale{1.08}
\plotone{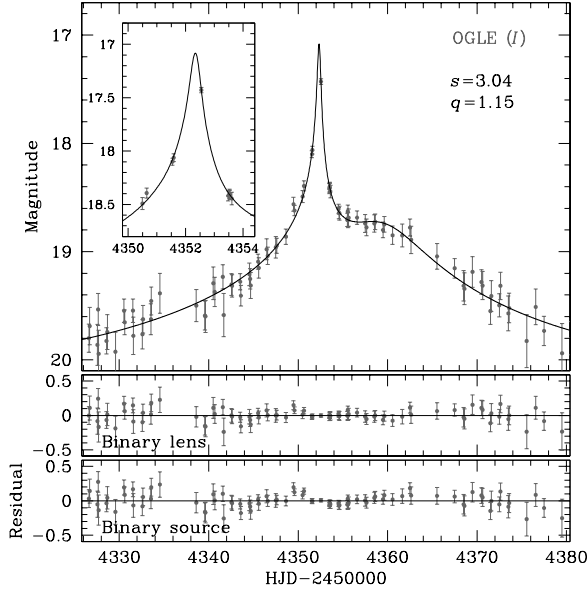}
\caption{\label{fig:five}
Light curve of OGLE-2007-BLG-491. Notations are same as in Fig 1. The two lower panels show the residuals from the binary-lens and binary-source models.
}\end{figure}

\begin{figure}[ht]
\epsscale{1.08}
\plotone{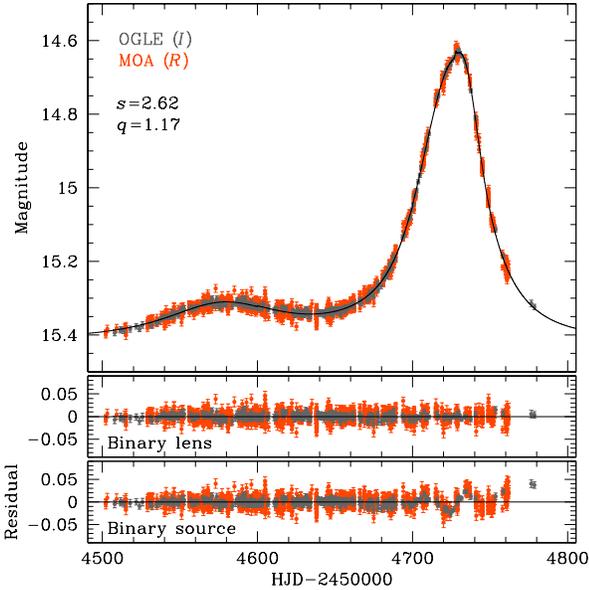}
\caption{\label{fig:six}
Light curve of OGLE-2008-BLG-143. The two lower panels show the residuals from the binary-lens and binary-source models. Due to the long-time gap between the two bumps at ${\rm HJD}\sim2454580$ and $\sim2454730$, lens parallax effect is detected.
}\end{figure}

\begin{figure}[ht]
\epsscale{1.08}
\plotone{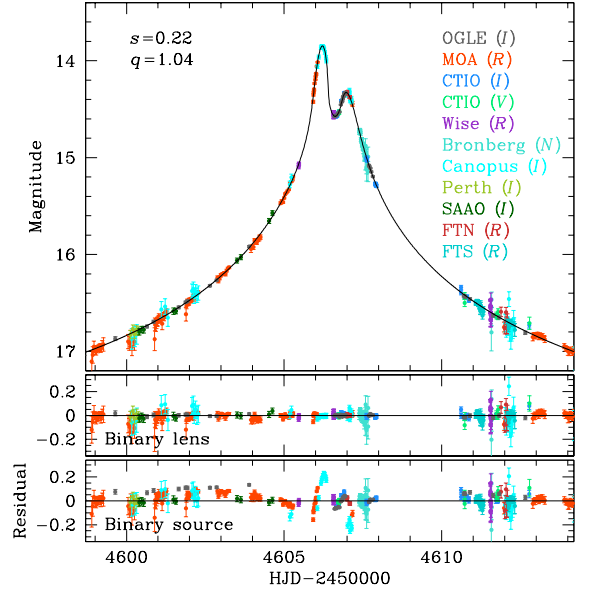}
\caption{\label{fig:seven}
Light curve of OGLE-2008-BLG-210. The two lower panels show the residuals from the binary-lens and binary-source models. With the data covering the two central bumps produced by caustic crossings, finite-source effects are detected.
}\end{figure}

\begin{figure}[ht]
\epsscale{1.08}
\plotone{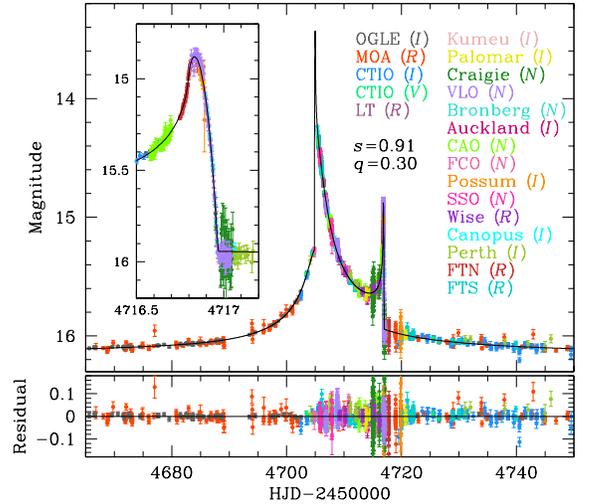}
\caption{\label{fig:eight}
Light curve of OGLE-2008-BLG-513. With dense coverage of the caustic exit by the additional data, finite-source effects are detected.
}\end{figure}

In Figures~\ref{fig:one} through \ref{fig:eight}, we present the light curves of the events. Also presented are the best-fit models obtained from our analyses and the residuals from the models. For the case where new solutions are identified, we present two panels of residuals from the new and previous solutions. See more details in Section 4.

Data sets used for the analyses were processed using the photometry codes developed by the individual groups. Since the individual data sets were obtained by using different telescopes, it is needed to readjust their error bars. The error bars of each data set is adjusted by
\begin{equation}
e'=k(e^2+e_{\rm min})^{1/2}.
\end{equation}
Here $e_{\rm min}$ is a term introduced so that the cumulative distribution function of $\chi^2$ as a function of lensing magnification becomes linear. This factor is needed to ensure that the dispersion of data points is consistent with error bars of the source brightness. The other term $k$ is a scaling factor used to make $\chi^2$ per degree of freedom (dof) becomes unity. This process is needed to ensure that each data set is fairly weighted according to its error bars.

\section{Analysis}

Light curves of binary-lens events are generally characterized by distinct sharp spikes occurring at the moments of caustic crossings. For events with such features (OGLE-2006-BLG-215, OGLE-2006-BLG-238, OGLE-2006-BLG-450, and OGLE-2008-BLG-513), we conduct binary-lens analyses. On the other hand, light curves of binary-lens events without caustic crossings do not exhibit such features and the resulting anomalies can often be imitated by those of binary-source events. For such events (OGLE-2007-BLG-159, OGLE-2007-BLG-491, OGLE-2008-BLG-143, and OGLE-2008-BLG-210), we conduct both binary-source and binary-lens analyses.

In the analyses, we consider higher-order effects. The first such an effect is caused by the finite size of the source star. For binary-lens events, this effect is important for caustic-crossing parts of the light curve where lensing magnifications vary abruptly with a small change of the source position and thus differential magnification on the surface of the source star becomes important. Among the analyzed binary-lens events, caustic crossings were resolved by the OGLE data for only one event (OGLE-2006-BLG-215), while caustics were resolved with combined data for all caustic-crossing events. We also consider the effects caused by the parallactic motion of the Earth and the orbital motion of lenses. These effects are important for long time-scale events where the duration of the event is equivalent to the orbital period of the Earth and lenses.

Modeling lensing events requires parameters that describe observed light curves. The light curve of a single-lens event is described by 3 parameters. They are the time of the closest approach of the source to the reference position of the lens, $t_0$, the lens-source separation at that moment, $u_0$ (impact parameter), and the event time scale, $t_{\rm E}$, which is defined as the time for the source to cross the Einstein radius of the lens. The Einstein radius is related to the physical parameters of the lens system by 
\begin{equation}
\theta_{\rm E}=(\kappa M\pi_{\rm rel})^{1/2};\:\:\pi_{\rm rel}={\rm AU}\left(\frac{1}{D_{\rm L}}-\frac{1}{D_{\rm S}}\right),
\end{equation}
where $\kappa=4G/\left(c^2{\rm AU}\right)$, {\it M} is the total mass of the lens, $\pi_{\rm rel}$ is the relative source-lens parallax, and $D_{\rm L}$ and $D_{\rm S}$ are the distances to the lens and source, respectively. The Einstein radius is used as a length scale in describing lensing phenomenon and $u_0$ is normalized by $\theta_{\rm E}$.

The light curve of a binary-source event corresponds to the linear sum of the fluxes from the two single-source events involved with the individual source stars. Then, describing the light curve requires two values of $t_0$ ($t_{0,1}$ and $t_{0,2}$) and $u_0$ ($u_{0,1}$ and $u_{0,2}$) but a single time scale because it is related only to the lens. One additional parameter is the flux ratio $q_{\rm F}$ between the source components. 

In order to model a binary-lens event, additional parameters are needed to describe the lens binarity. They are the projected separation $s$ and the mass ratio $q$ between the lens components. We note that the separation is normalized by $\theta_{\rm E}$. Due to the lens binarity, positions around the lens are no longer radially symmetric. Then, one needs an additional parameter $\alpha$ designating the angle between the source trajectory and the line connecting the binary lens components (source trajectory angle).

\begin{deluxetable*}{lclllllllll}{ht}
\tablecaption{Binary lens parameters of newly analyzed events\label{table:three}}
\tablewidth{0pt}
\tablehead{
\multicolumn{1}{c}{event} &
\multicolumn{1}{c}{solution} &
\multicolumn{1}{c}{$\chi^2/{\rm dof}$} &
\multicolumn{1}{c}{$u_0$} &
\multicolumn{1}{c}{$t_{\rm E}$} &
\multicolumn{1}{c}{$s$} &
\multicolumn{1}{c}{$q$} &
\multicolumn{1}{c}{$\alpha$} &
\multicolumn{1}{c}{$\rho$} &
\multicolumn{1}{c}{$\pi_{{\rm E},N}$} &
\multicolumn{1}{c}{$\pi_{{\rm E},E}$} \\
\multicolumn{1}{c}{} &
\multicolumn{1}{c}{} &
\multicolumn{1}{c}{} &
\multicolumn{1}{c}{} &
\multicolumn{1}{c}{(days)} &
\multicolumn{1}{c}{} &
\multicolumn{1}{c}{} &
\multicolumn{1}{c}{(rad)} &
\multicolumn{1}{c}{$(10^{-3})$} &
\multicolumn{1}{c}{} &
\multicolumn{1}{c}{}
}
\startdata
OGLE-2006-BLG-215 &         & 1628.3 & 0.468      & 27.4     & 1.01      & 1.09      & 5.828      &  9.07     & -- & -- \\
                  &         & /1552  & $\pm$0.008 & $\pm$0.4 & $\pm$0.01 & $\pm$0.05 & $\pm$0.009 & $\pm$0.21 & -- & -- \\
\hline
OGLE-2006-BLG-238/& $s<1$   & 2177.6 & 0.079      & 14.9     & 0.90      & 0.08      & 1.687      & 1.86      & -- & -- \\
MOA-2006-BLG-26   &         & /2151  & $\pm$0.008 & $\pm$0.8 & $\pm$0.01 & $\pm$0.01 & $\pm$0.023 & $\pm$0.26 & -- & -- \\
                  & $s>1$   & 2163.7 & 0.068      & 15.3     & 1.33      & 0.12      & 1.651      & 2.44      & -- & -- \\
                  &         & /2151  & $\pm$0.008 & $\pm$0.8 & $\pm$0.02 & $\pm$0.01 & $\pm$0.010 & $\pm$0.35 & -- & -- \\
\hline
OGLE-2006-BLG-450 &         & 1008.8 & 0.139      & 8.6      & 0.90      & 0.59      & 1.258      & 10.31     & -- & -- \\ 
                  &         & /1007  & $\pm$0.012 & $\pm$0.2 & $\pm$0.01 & $\pm$0.03 & $\pm$0.021 & $\pm$0.41 & -- & -- \\ 
\hline
OGLE-2007-BLG-159 &         & 1634.7 & 0.684      & 18.0     & 1.10      & 1.61      & -0.331     & -- & -- & -- \\ 
                  &         & /1639  & $\pm$0.008 & $\pm$0.2 & $\pm$0.01 & $\pm$0.05 & $\pm$0.004 & -- & -- & -- \\ 
\hline
OGLE-2007-BLG-491 & $s<1$   & 910.2  & 0.244      & 32.1     & 0.61      & 0.35      & 3.774      & -- & -- & -- \\ 
                  &         & /904   & $\pm$0.035 & $\pm$3.2 & $\pm$0.03 & $\pm$0.08 & $\pm$0.073 & -- & -- & -- \\ 
                  & $s>1$   & 907.0  & 0.090      & 61.0     & 3.04      & 1.15      & 3.483      & -- & -- & -- \\ 
                  &         & /904   & $\pm$0.019 & $\pm$5.6 & $\pm$0.35 & $\pm$0.35 & $\pm$0.053 & -- & -- & -- \\ 
\hline
OGLE-2008-BLG-143/& $u_0>0$ & 4598.8 & 0.191      & 62.9     & 2.62      & 1.17      & 2.566      & 15.34     & 0.19      & 0.16      \\ 
MOA-2008-BLG-111  &         & /4786  & $\pm$0.002 & $\pm$0.4 & $\pm$0.02 & $\pm$0.05 & $\pm$0.005 & $\pm$1.16 & $\pm$0.01 & $\pm$0.01 \\ 
                  & $u_0<0$ & 4606.9 & -0.202     & 60.5     & 2.70      & 1.31      & -2.578     & 16.11     & -0.12     & 0.18      \\ 
                  &         & /4786  & $\pm$0.002 & $\pm$0.5 & $\pm$0.02 & $\pm$0.04 & $\pm$0.005 & $\pm$1.65 & $\pm$0.01 & $\pm$0.02 \\ 
\hline
OGLE-2008-BLG-210/&         & 3436.2 & 0.027      & 19.3     & 0.22      & 1.04      & 5.265      & 12.32     & -- & -- \\ 
MOA-2008-BLG-177  &         & /3480  & $\pm$0.001 & $\pm$0.2 & $\pm$0.01 & $\pm$0.05 & $\pm$0.006 & $\pm$0.24 & -- & -- \\ 
\hline
OGLE-2008-BLG-513/&         & 4951.7 & 0.079      & 35.4     & 0.91      & 0.30      & 5.590      & 2.38      & -- & -- \\ 
MOA-2008-BLG-401  &         & /4972  & $\pm$0.001 & $\pm$0.5 & $\pm$0.01 & $\pm$0.01 & $\pm$0.003 & $\pm$0.03 & -- & -- 
\enddata
\tablecomments{
For OGLE-2008-BLG-143, two solutions resulting from the ecliptic degeneracy (with $u_0>0$ and $u_0<0$) are presented. For each of OGLE-2006-BLG-238, OGLE-2007-BLG-491, and OGLE-2008-BLG-210, two solutions resulting from the close/wide degeneracy (with $s<1$ and $s>1$) are presented.
}
\end{deluxetable*}


In order to consider higher-order effects, one also needs additional parameters. Accounting for finite-source effects requires a parameter defined as the angular source radius $\theta_\star$ expressed in units of the Einstein radius $\theta_{\rm E}$, $\rho =\theta_\star/\theta_{\rm E}$ (normalized source radius). Considering parallax effects requires two parameters $\pi_{{\rm E},N}$ and $\pi_{{\rm E},E}$ that are the north and east components of the lens parallax vector, $\pivec_{\rm E}$, projected on the sky in the north and east equatorial coordinates, respectively. The lens parallax vector is defined as
\begin{equation}
\pivec_{\rm E}=\frac{\pi_{\rm rel}}{\theta_{\rm E}}\frac{\muvec}{\mu},
\end{equation}
 where $\muvec$ represents the vector of the relative lens-source proper motion. In order to account for lens orbital effects, one needs two parameters including the change rates of the projected binary separation, $ds/dt$, and the source trajectory angle, $d\alpha/dt$. When either the parallax or the lens-orbital effects are considered, we test two solutions with the lens-source impact parameters $u_0>0$ and $u_0<0$ in order to check the known ``ecliptic degeneracy'' \citep{skowron11}.

In our analysis, we search for the set of the lensing parameters that best describe the observed light curves. For binary-source modeling where lensing magnifications vary smoothly with the parameters, searching for lensing solutions is done by minimizing $\chi^2$ using a downhill approach. For the downhill approach, we use the Markov Chain Monte Carlo (MCMC) method. For binary-lens modeling, on the other hand, the search is done through multiple steps. In the first step, we inspect local minima by exploring $\chi^2$ surface in the parameter space. For this, we conduct a grid search for a subset of the lensing parameters. We choose $(s,q,\alpha)$ as grid parameters because lensing magnifications can vary dramatically with the small changes of these parameters. On the other hand, lensing magnifications vary smoothly with the changes of the other parameters and thus we search for these parameters by using a downhill approach. In the second step, we refine each local minimum by letting all lensing parameters vary around the minimum. In the final step, we find the global minimum by comparing $\chi^2$ values of the identified local minima. This multiple-step process allows one to check the existence of possible degenerate solutions that cause confusion in the interpretation of the observed light curves. 

In computing finite-source magnifications, we consider the limb-darkening variation of source stars. For this, we model the surface brightness profile of the source star as
\begin{equation}
S_\lambda \propto 1-\Gamma_{\lambda}(1-1.5\cos\psi),
\end{equation}
where $\Gamma_\lambda$ is the limb-darkening coefficient and $\psi$ is the angle between the line of sight toward the source center and the normal to the source surface. We adopt the limb-darkening coefficients from \citet{claret00} based on the source types that are determined from the de-reddened color and brightness of the source stars measured by using the multi-band ({\it V} and {\it I}) data taken from CTIO observation. Among the 6 caustic-crossing events, there exist multi-band data for 3 events and thus the source types are able to be determined. For OGLE-2006-BLG-215 and OGLE-2006-BLG-450, which were observed in {\it I} band only, we adopt $\Gamma_I=0.5$ by taking a median value of Bulge stars.

\begin{figure*}[ht]
\epsscale{0.75}
\plotone{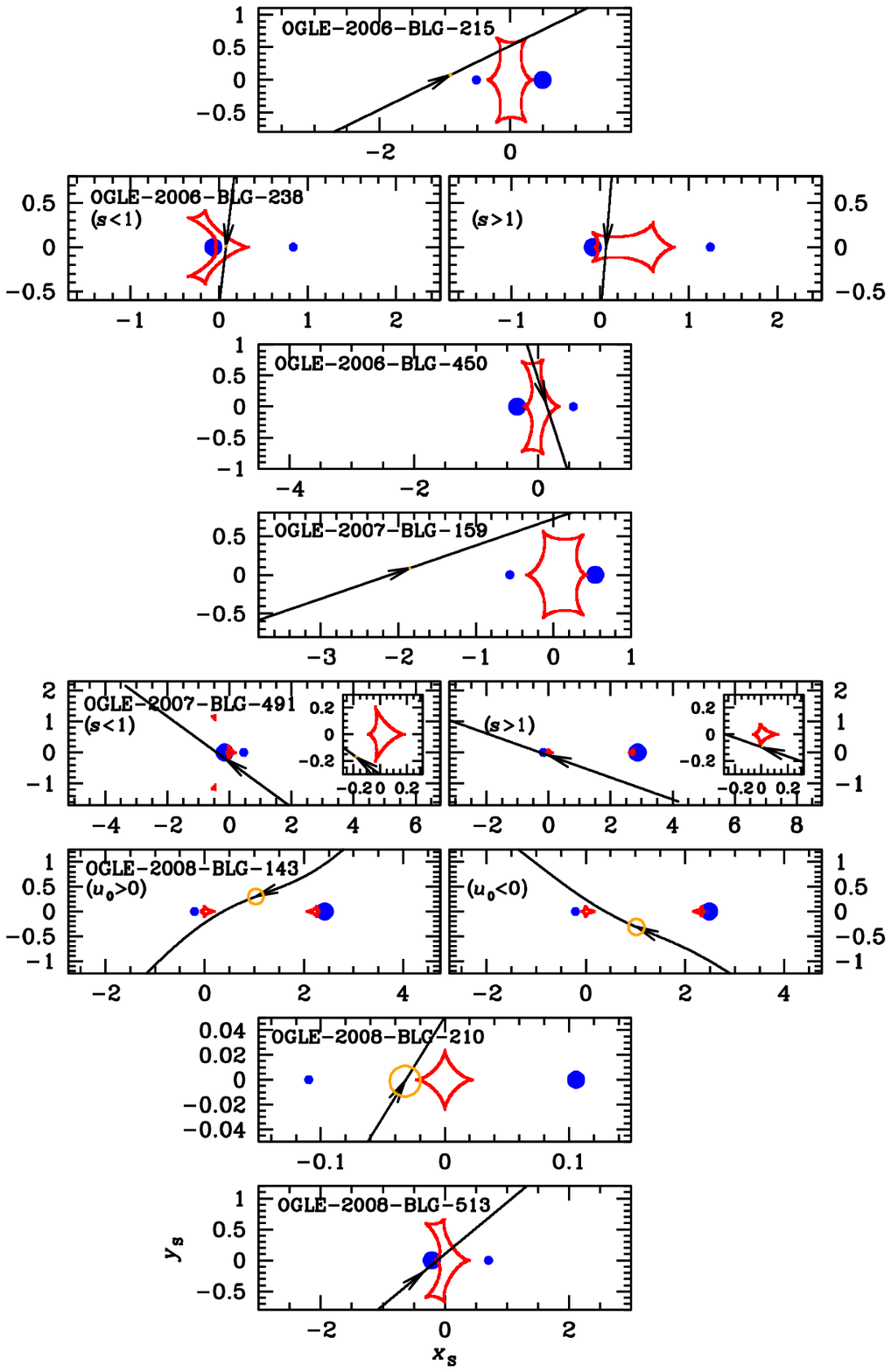}
\caption{\label{fig:nine}
Caustic structures and source trajectories for the lensing solutions of the binary-lens events presented in Fig.\ \ref{fig:one} through \ref{fig:eight}. In each panel, the two filled dots represent the positions of the binary lens components where the bigger dot is the higher-mass component. The closed curves with cusps represent the caustics. The curve with an arrow is the source trajectory. For events with identified degenerate solutions, we present two sets of the lens-system geometry corresponding to the individual solutions. All lengths are scaled by the Einstein radius corresponding to the total mass of the lens.
}\end{figure*}

\section{Result}

In Table~\ref{table:three}, we present the best-fit lensing parameters obtained from modeling. In order to present lensing parameters of all events in a single table, we do not present the value of $t_0$ that has no physical meaning in describing a lens system. It is found that all 8 analyzed events are interpreted to be caused by binary lenses. The model light curves corresponding to the best-fit solutions of the individual events are superposed on the light curves in Figures~\ref{fig:one} through \ref{fig:eight}. 

We find that the previous binary-source interpretations of 5 events are better interpreted by binary-lens models. These events include OGLE-2006-BLG-238, OGLE-2007-BLG-159, OGLE-2007-BLG-491, OGLE-2008-BLG-143, and OGLE-2008-BLG-210.

With additional data covering caustic crossings, we additionally detect finite-source effects for 6 events. These events include OGLE-2006-BLG-215, OGLE-2006-BLG-238, OGLE-2006-BLG-450, OGLE-2008-BLG-143, OGLE-2008-BLG-210, and OGLE-2008-BLG-513. Among them, we are able to measure the Einstein radii of 3 events for which multi-band data are available and thus source types are known from color information. These events are OGLE-2006-BLG-238, OGLE-2008-BLG-210, and OGLE-2008-BLG-513. The Einstein radius is measured by $\theta_{\rm E}=\theta_\star/\rho$, where the angular source radius $\theta_\star$ is estimated from the de-reddened color and brightness of the source star and the normalized source radius $\rho$ is measured from modeling the light curve. The source radius is determined following the method of \citet{yoo04}, where we first locate the source star on the color-magnitude diagram of stars in the field and then calibrate the de-reddened color and brightness of the source star by using the centroid of the red clump giants as a reference under the assumption that the source and red clump giants experience the same amount of extinction and reddening. 

From the inspection of additional higher-order effects, we detect clear signature of parallax effects for OGLE-2008-BLG-143. For OGLE-2006-BLG-215, we initially detect both parallax and lens-orbital effects from the analysis based on the online data, but find that the signals of these higher-order effects are spurious from the analysis based on the re-reduced data. 

Finally, we identify additional solutions caused by the known degeneracy for 3 events. For OGLE-2006-BLG-238 and OGLE-2007-BLG-491, we identify degenerate solutions caused by the ``close/wide'' degeneracy. This degeneracy is caused by the symmetry of the lens-mapping equations between the binary lenses with separations $s$ and $s^{-1}$ \citep{griest98,dominik99,an05}. For OGLE-2008-BLG-143, we identify degenerate solutions caused by the ecliptic degeneracy. This degeneracy is caused by the mirror symmetry between the source trajectories with $u_{0}>0$ and $u_{0}<0$ with respect to the binary axis.

Below we describe details of the results from our analyses of the individual events. We then compare the results with those from the previous analyses. 

\subsection{OGLE-2006-BLG-215}

We find that the event was produced by the crossings of a source star over a single big caustic produced by a binary lens with a projected separation similar to the Einstein radius, i.e. $s\sim1$ (resonant binary). For typical caustic-crossing events, the light curve within the two successive caustic crossings has a ``U'' shape, but the light curve of this event exhibits deviations from this shape. We find that the deviation was produced by the asymptotic approach of the source trajectory to the caustic line. See the source trajectory and caustic configuration presented in Figure~\ref{fig:nine}. The previous interpretation of the event is basically same as our interpretation.

From the initial analysis based on the online data, we find that the light curve is affected by both parallax and lens-orbital effects. From additional analysis based on the re-reduced data, however, we find that the signature of these higher-order effects is false positive caused by the poor photometry, suggesting the need for careful analysis in detecting subtle higher-order effects. In Figure~\ref{fig:one}, we present the residuals from the online and re-reduced data.

The caustic-crossing part of the light curve was densely resolved and thus the normalized source radius is precisely measured. However, there exists no multi-band data required to estimate the angular source radius and thus the Einstein radius cannot be measured.

\subsection{OGLE-2006-BLG-238}

In addition to the OGLE group, the event was observed by many other groups and the MOA group designated the event MOA-2006-BLG-26. In the OGLE Anomaly Catalog, this event was interpreted as a binary-source event. With the addition of follow-up data, especially the SAAO data set covering the caustic exit, it is clear that the event was produced by a binary lens. We find that a binary-lens interpretation is better than the binary-source interpretation by $\Delta\chi^2=5233.4$. See the residuals from the binary-lens and binary-source models presented in the lower two panels of Figure~\ref{fig:two}. From modeling the light curve including additional follow-up data, it is found that the event was produced by the source crossings over the single big caustic of a resonant binary. See Figure~\ref{fig:nine} for the source trajectory with respect to the caustic.

With the coverage of the caustic crossing by the SAAO data and multi-band data obtained from CTIO observation, we measure the Einstein radius of the lens system. The measured de-reddened color and brightness of the source stars are $(V-I,I)_{0}=(1.24,17.7)$, which correspond to a K-type Bulge subgiant with $\theta_\star=1.66\pm0.14$ $\mu$-arcsec. With the measured normalized source radius $\rho=(2.44\pm0.35)\times10^{-3}$, then, the Einstein radius is $\theta_{\rm E}=0.68\pm0.12$ milli-arcsec.  

We find that there exist two local solutions resulting from the close/wide degeneracy. The degeneracy is quite severe and the wide-binary solution $(s>1)$ is preferred over the close-binary solution $(s<1)$ by merely $\Delta\chi^2=13.9$. We present the lensing parameters of both close and wide binary solutions and the corresponding lens-system geometry in Table~\ref{table:three} and Figure~\ref{fig:nine}, respectively. Due to the short time scale of the event, $t_{\rm E}\sim15.3$ days, neither parallax nor lens-orbital effect is measured. 

\subsection{OGLE-2006-BLG-450}

It is found that this event was also produced by a binary lens with a resonant separation similar to the previous two events. See Figure~\ref{fig:nine} for the source trajectory with respect to the caustic. The interpretation is basically same as the previous one. Compared to the previous analysis, we additionally measure the normalized source radius because the caustic exit of the event was resolved by the data from Canopus observation. See Figure~\ref{fig:three}. However, there exists no multi-band data and thus the Einstein radius cannot be measured. 

\subsection{OGLE-2007-BLG-159}

The event was interpreted as a binary-source event in the previous analysis. From our analysis with additional follow-up data that cover the bump at ${\rm HJD}\sim2454227$ and the falling part of the light curve, we find that the light curve is better interpreted by a binary-lens event. The $\chi^2$ difference between the binary-lens and binary-source models is $\Delta\chi^2=611.6$. See also the residuals from the two models presented in the lower two panels of Figure~\ref{fig:four}.

We find that the event was produced by the source approach to a cusp of a big caustic produced by a resonant binary. See Figure~\ref{fig:nine} for the caustic configuration and the source trajectory. Since the source did not cross the caustic, finite-source effects are not detected. The event time scale is short, $t_{\rm E}\sim18.0$ days, and thus parallax effect is not detected either. 

\begin{deluxetable*}{lccc}
\centering
\tablecaption{Ranges of the lens mass and distance\label{table:four}}
\tablewidth{0pt}
\tablehead{
\multicolumn{1}{c}{Event} &
\multicolumn{2}{c}{Mass ($M_{\odot}$)} &
\multicolumn{1}{c}{Distance} \\
\multicolumn{1}{c}{} &
\multicolumn{1}{c}{primary} &
\multicolumn{1}{c}{companion} &
\multicolumn{1}{c}{(kpc)}
}
\startdata
OGLE-2006-BLG-215                   & 0.38$\pm$0.23 & 0.11$\pm$0.07 & 6.36$\pm$1.52 \\
OGLE-2006-BLG-238/MOA-2006-BLG-26   & 0.41$\pm$0.22 & 0.11$\pm$0.06 & 4.81$\pm$1.09 \\
OGLE-2006-BLG-450                   & 0.19$\pm$0.13 & 0.03$\pm$0.02 & 7.28$\pm$1.25 \\
OGLE-2007-BLG-159                   & 0.29$\pm$0.19 & 0.07$\pm$0.05 & 6.71$\pm$1.40 \\
OGLE-2007-BLG-491                   & 0.49$\pm$0.28 & 0.18$\pm$0.10 & 5.63$\pm$1.98 \\
OGLE-2008-BLG-143/MOA-2008-BLG-111  & 0.36$\pm$0.19 & 0.30$\pm$0.16 & 6.78$\pm$0.90 \\
OGLE-2008-BLG-210/MOA-2008-BLG-177  & 0.26$\pm$0.17 & 0.05$\pm$0.04 & 7.35$\pm$0.93 \\
OGLE-2008-BLG-513/MOA-2008-BLG-401  & 0.45$\pm$0.19 & 0.10$\pm$0.04 & 6.98$\pm$0.93 
\enddata
\end{deluxetable*}

\subsection{OGLE-2007-BLG-491}

The light curve of the event is characterized by two bumps: a strong bump at ${\rm HJD}\sim2454352$ and a weak bump at ${\rm HJD}\sim2454360$. The event was previously interpreted as a binary-source event. From our analysis, however, it is found that the event is better interpreted by a binary-lens model. The $\chi^2$ difference between the binary-lens and binary-source models is $\Delta\chi^2=35.3$. See the residual from the two models in Figure~\ref{fig:five}. According to the binary-lens model, the strong bump was produced by the source crossing over the tip of a small caustic and the weak bump was produced by the source passage around another cusp of the caustic. See Figure~\ref{fig:nine}.

We find a pair of degenerate solutions caused by the close/wide degeneracy. The degeneracy is severe with $\Delta\chi^2=3.2$. We present the lensing parameters and the configurations of the lens system of the two degenerate solutions in Table~\ref{table:three} and Figure~\ref{fig:nine}, respectively. Although the source trajectory crossed the tip of the caustic, the observational cadence during the caustic crossing is not high enough to yield reliable detection of finite-source effects. Other higher-order effects are not detected either.

\subsection{OGLE-2008-BLG-143}

The event was discovered and monitored not only by the OGLE survey but also by the MOA survey. The event was dubbed as MOA-2008-BLG-111 by the MOA group.  

The light curve of the event is characterized by two bumps that are separated by $\sim150$ days. Previous analysis interpreted that the event was produced by a binary source. However, we find that the binary-lens interpretation better fits the light curve than the binary-source interpretation with $\Delta\chi^2=617.1$. See the residual from the two models presented in Figure~\ref{fig:six}. 

According to the best-fit binary-lens model, the event was produced by a wide binary with a projected separation between the lens components $s\sim2.6$ and the two bumps were produced by the successive approaches of the source to the individual lens components. See Figure~\ref{fig:nine} for the source trajectory with respect to the lens system.

Due to the long time gap between the two bumps, it is possible to measure higher-order effects. We find that considering the parallax effect improves the fit by $\Delta\chi^2=254.6$. On the other hand, the effect of the lens orbital motion is negligible due to the wide separation between the lens components and thus long orbital period of the binary. The source crossed the tip of the caustic during the second caustic approach and thus finite-source effects are detected. However, no multi-band observation was conducted and thus the Einstein radius cannot be measured. 

It is found that the ecliptic degeneracy is severe with $\Delta\chi^2=8.1$. We present both $u_0>0$ and $u_0<0$ solutions in Table~\ref{table:three}. The source trajectories and caustic configurations corresponding to the individual solutions are presented in Figure~\ref{fig:nine}.

\subsection{OGLE-2008-BLG-210}

The light curve of the event is characterized by two short-term bumps that occurred near the peak. The anomaly is partially covered by the OGLE data but it is substantially better covered with additional follow-up data. The MOA group also detected the event and dubbed it MOA-2008-BLG-177. The event was interpreted as a binary-source event in the previous analysis based on only the OGLE data. 

By conducting both binary-source and binary-lens modeling with combined data, we find that the light curve is better explained by a binary-lens interpretation. The $\chi^2$ difference between the two models is $\Delta\chi^2=6045.6$. See the residuals from the individual models presented in Figure~\ref{fig:seven}.

We find two local solutions resulting from the close/wide degeneracy. The degeneracy is severe with $\Delta\chi^2=32.5$. For the close-binary solution, the lensing parameters are well defined with $s=0.22$ and $q=1.04$. On the other hand, the lensing parameters of the wide-binary solution are very uncertain because the lens is in the Chang-Refsdal regime $(s\gg1)$, where continuous degeneracy with different combinations of $s$ and $q$ exists \citep{kim08}. Due to large uncertainties of the wide-binary parameters, we present only the solution of the close-binary model in Table~\ref{table:three}. For the binary with a separation $s\ll1$ or $s\gg1$, the resulting caustic has an astroid shape with 4 cusps. The central anomaly in the light curve was produced by the source crossings over the two cusps of the astroid-shape caustic. See Figure~\ref{fig:nine}. 

Since the source crossed the caustic and the caustic crossing was densely resolved, finite-source effects are clearly detected. Combined with the de-reddened color and brightness of the source star $(V-I,I)_0=(1.09,16.3)$ determined based on the multi-band data taken from CTIO observation, we estimate that the angular sourc radius $\theta_\star=2.67\pm0.19$ $\mu$-arcsec. Combined with the normalized source radius $\rho$, we estimate the angular Einstein radius $\theta_{\rm E}=0.22\pm0.02$ milli-arcsec. Neither parallax nor lens orbital effect is detected due to the short time scale, $t_{\rm E}\sim19.3$ days, of the event. 

\subsection{OGLE-2008-BLG-513}

The light curve of the event exhibits obvious caustic-crossing features at ${\rm HJD}\sim2454705$ and $\sim2454717$. Besides the OGLE data, the event was observed from 18 observatories including the MOA survey and these additional data densely cover the caustic exit. The event was dubbed as MOA-2008-BLG-401 in the MOA event list. Our analysis of the light curve is consistent with the previous one where the event was produced by the source crossing over the caustic of a resonant binary.  

With additional data covering the caustic exit, finite-source effects are clearly detected and the normalized source radius is precisely measured. With multi-band data from CTIO observation, we estimate the angular source radius as $\theta_\star=0.88\pm0.06$ $\mu$-arcsec based on the de-reddened color and magnitude $(V-I,I)_0=(0.16,16.7)$, resulting in the measured Einstein radius $\theta_{\rm E}=0.37\pm0.03$ milli-arcsec. We note that the measured source color is very atypical for Bulge stars. This indicates the possibility that the source might be in the Disk not in the Bulge. If this is the case, the de-reddened and brightness cannot be estimated based on the Bulge clump giants because the source and Bulge giants experience different amount of reddening and extinction.

Modeling including parallax effects yields a slightly improved fit to the overall data. However, the improvement is marginal with $\Delta\chi^2\sim31$. Furthermore, while the parallax model provides a better fit to the MOA data, it provides a poorer fit to the OGLE data. Therefore, we judge that the marginal improvement by the parallax model is mostly ascribed to the systematics of data. 

\subsection{Physical Lens Parameters}

Among the analyzed events, there is no event for which both the Einstein radius and the lens parallax are simultaneously measured and thus the mass and distance to the lens are uniquely measured. Nevertheless, it is still possible to constrain the physical lens parameters based on the event time scale. For events where either the Einstein radius or the lens parallax is additionally measured, the physical parameters can be better constrained.

To estimate the ranges of the mass and distance to each lens, we conduct Bayesian analysis based on the mass function, and spatial and velocity distributions of Galactic stars \citep{han95}. In Table~\ref{table:four}, we present the estimated ranges of the physical lens parameters. To be noted is that the masses of the binary companions of the lens for the events OGLE-2006-BLG-450, OGLE-2008-BLG-210 are in the brown-dwarf regime. For OGLE-2007-BLG-159, the companion mass is at the star/brown-dwarf boundary.

\section{Conclusion}
We reanalyzed anomalous lensing events in the OGLE-III EWS database. With the addition of data from other survey and follow-up observations, we conducted thorough search for possible degenerate solutions and investigated higher-order effects. Among the analyzed events, we presented analyses of 8 events for which either new solutions were identified by resolving degeneracy in lensing solutions or additional information was obtained by detecting higher-order effects. We found that the previous binary-source interpretations of 5 events (OGLE-2006-BLG-238, OGLE-2007-BLG-159, OGLE-2007-BLG-491, OGLE-2008-BLG-143, and OGLE-2008-BLG-210) were better interpreted by binary-lens models. In addition, we detected finite-source effects for 6 events (OGLE-2006-BLG-215, OGLE-2006-BLG-238, OGLE-2006-BLG-450, OGLE-2008-BLG-143, OGLE-2008-BLG-210, and OGLE-2008-BLG-513) with additional data covering caustic crossings and measure the Einstein radii for 3 events (OGLE-2006-BLG-238, OGLE-2008-BLG-210, and OGLE-2008-BLG-513). For OGLE-2008-BLG-143, we detect clear signature of parallax effects and measure the lens parallax. In addition, we presented degenerate solutions resulting from the known close/wide or ecliptic degeneracy. Finally, we note that the masses of the binary companions of the lenses of OGLE-2006-BLG-450 and OGLE-2008-BLG-210 are likely in the brown-dwarf regime according to the mass ranges estimated by Bayesian analysis.

\mbox{}

\acknowledgments 
Work by C.H. was supported by the Creative Research Initiative Program (2009-0081561) of the National Research Foundation of Korea. The OGLE project has received funding from the European Research Council under the European Community's Seventh Framework Programme (FP7/2007-2013) / ERC grant agreement no.~246678 to AU. The MOA project is supported by the grants JSPS18253002 and JSPS20340052. T.S. acknowledges the financial support from the JSPS, JSPS23340044, JSPS24253004. D.P.B. was supported by grants NASA-NNX12AF54G, JPL-RSA 1453175 and NSF AST-1211875. B.S.G. and A.G. were supported by NSF grant AST 110347. B.S.G., A.G., R.P.G. were supported by NASA grant NNX12AB99G. Work by J.C.Y. was performed in part under contract with the California Institute of Technology (Caltech) funded by NASA through the Sagan Fellowship Program.

\clearpage

\end{document}